\pdfoutput=1
\documentclass[aps, prl,twocolumn,showpacs,preprintnumbers,superscriptaddress,amsmath,amssymb]{revtex4-2}
\usepackage{graphicx}
\usepackage{float}
\usepackage{dcolumn}
\usepackage{color}
\usepackage[dvipsnames]{xcolor} 
\usepackage{latexsym,bm}
\usepackage[normalem]{ulem}
\usepackage{multirow}
\usepackage{appendix}
\usepackage{amsmath}
\usepackage{amsfonts}
\usepackage{booktabs}
\usepackage{subcaption}
\usepackage{tabularx}

\usepackage{threeparttable} 

\newcommand{\SL}[1]{\textcolor{black}{{#1}}}

\begin{document}

\hyphenpenalty=5000
\tolerance=1000

\title{Extending Nonlocal Kinetic Energy Density Functionals to Isolated Systems via a Density-Functional-Dependent Kernel}

\author{Liang Sun}
\affiliation{HEDPS, CAPT, School of Physics and School of Mechanics and Engineering Science, Peking University, Beijing 100871, P. R. China}
\author{Mohan Chen}
\email{mohanchen@pku.edu.cn}
\affiliation{HEDPS, CAPT, School of Physics and School of Mechanics and Engineering Science, Peking University, Beijing 100871, P. R. China}
\date{\today}

\begin{abstract}
The Wang-Teter-like nonlocal kinetic energy density functional (KEDF) in the framework of orbital-free density functional theory, while successful in some bulk systems, exhibits a critical Blanc-Canc\`es instability [J. Chem. Phys. \textbf{122}, 214106 (2005)] when applied to isolated systems, where the total energy becomes unbounded from below.
We trace this instability to the use of an ill-defined average charge density, which causes the functional to simultaneously violate the scaling law and the positivity of the Pauli energy.
By rigorously constructing a density-functional-dependent kernel, we resolve these pathologies while preserving the formal exactness of the original framework.
By systematically benchmarking single-atom systems of 56 elements, we find the resulting KEDF retains computational efficiency while achieving an order-of-magnitude accuracy enhancement over the WT KEDF. In addition, the new KEDF preserves WT's superior accuracy in bulk metals, outperforming the semilocal functionals in both regimes.
\end{abstract}
\maketitle

\section{Introduction}

Orbital-free density functional theory (OFDFT)~\cite{02Carter, 23CR-Mi} circumvents the $O(N^3)$ ($N$ is the electron number) computational bottleneck of Kohn-Sham DFT (KSDFT)~\cite{64PR-Hohenberg, 65PR-Kohn-ksdft} by eliminating explicit orbital dependence, achieving near-linear scaling efficiency.
The method has demonstrated promising applications in diverse systems including liquid metal surface structures~\cite{04PRL-Gonzalez-liquid}, lightweight alloy design~\cite{14AM-Shin}, and warm dense matter simulations~\cite{13PRL-White-wdm, 14PRL-Sjostrom-wdm, 18L-Ding-oftddft}.
However, current OFDFT still encounters enormous challenges in approximating the non-interacting kinetic energy $T_s[\rho]$ solely through the electron density $\rho(\mathbf{r})$ --- a fundamental conundrum posed by the Hohenberg-Kohn theorem~\cite{64PR-Hohenberg, 11CTC-Perdew}.  
Current kinetic energy density functionals (KEDFs) either sacrifice accuracy for universality or fail to maintain numerical stability in finite systems, severely restricting OFDFT's applicability beyond bulk materials.  

Semilocal KEDFs, constructed from the electron density $\rho(\mathbf{r})$, its dimensionless gradient $s=|\nabla\rho|/(2k_{\mathrm{F}}\rho)$ and Laplacian $q=\nabla^2\rho/(4k_{\mathrm{F}}^2\rho)$ (where $k_{\mathrm{F}}=(3\pi^2\rho)^{1/3}$), maintain universality across bulk and isolated systems through their locally defined variables.~\cite{27-Thomas-local, 27TANL-Fermi-local, 35-vW-semilocal, 57SPJETP-Kirzhnits-GE2, 07PRB-Perdew-semilocal-mgga, 11PRL-Constantin-semilocal, 18B-Luo-semilocal, 18JPCL-Constantin-semilocal, 24JCTC-Wang-semilocal}
Their neglect of nonlocal information, however, leads to systematic errors in key electronic properties: atomic shell structures vanish~\cite{02Carter}, charge density cusps at nuclei are unphysical unless additional correction introduced~\cite{57CPAM-Kato-cusp, 24B-Andrew-cusp}.

Nonlocal KEDFs enhance accuracy by incorporating nonlocal information through convolution kernels $w(\mathbf{r},\mathbf{r}^{\prime})$.  
The Wang-Teter-like (WT-like) functionals, including WT~\cite{92B-Wang-nonlocal}, Smargiassi-Madden (SM)~\cite{94B-Smargiassi-nonlocal}, and Perrot~\cite{94JPCM-Perrot-nonlocal}, employ a density-independent kernel $w(k_{\mathrm{F}}^0,|\mathbf{r}-\mathbf{r}^{\prime}|)$, parametrized by $k_{\mathrm{F}}^0=(3\pi^2\rho_0)^{1/3}$.
Here, $\rho_0$ is a parameter conventionally set as the average charge density $\rho_\mathrm{avg}$.
The WT-like KEDFs generally decompose the non-interacting kinetic energy $T_s$ as
\begin{equation}
    T_{s} = T_{\mathrm{TF}} + T_{\mathrm{vW}} + T_{\mathrm{NL}},
\end{equation}
where $T_{\mathrm{TF}}=C_{\mathrm{TF}}\int{\rho^{5/3}(\mathbf{r}){\mathrm{d}}^3 \mathbf{r}}$ represents the Thomas-Fermi (TF) KEDF with $C_{\mathrm{TF}}=\frac{3}{10}(3\pi^2)^{2/3}$,~\cite{27-Thomas-local, 27TANL-Fermi-local} $T_{\mathrm{vW}}=\frac{1}{2}\int{\left|\nabla\sqrt{\rho(\mathbf{r})}\right|^2{\mathrm{d}}^3 \mathbf{r}}$ is the von Weizs$\mathrm{\Ddot{a}}$cker (vW) KEDF serving as a rigorous lower bound to $T_s$,~\cite{35-vW-semilocal} and $T_{\mathrm{NL}}$ encodes nonlocal information through the following convolution
\begin{equation}
    \begin{aligned}
        T_{\mathrm{NL}} &= C_{\mathrm{TF}} \iint{\rho^{\alpha}(\mathbf{r}) w(k_{\mathrm{F}}^0,\mathbf{r}-\mathbf{r^{\prime}}) \rho^{\beta}(\mathbf{r^{\prime}}){\mathrm{d}}^3 \mathbf{r}{\mathrm{d}}^3 \mathbf{r^\prime}}.
    \end{aligned}
\end{equation}
%
%
Distinct physical regimes motivate specific parameterizations: The original WT KEDF ($\alpha$=$\beta$=$\frac{5}{6}$) targets weakly varying uniform electron gas (UEG), while the Perrot KEDF ($\alpha$=$\beta$=1) was constructed for thin electron gas, and the SM KEDF ($\alpha$=$\beta$=$\frac{1}{2}$) is derived from low-$q$ limit.
Despite achieving $O(N\ln N)$ scaling and improved accuracy in bulk systems, these functionals catastrophically failure in isolated systems due to the Blanc-Canc\`es (BC) instability~\cite{05JCP-Blanc-BC}, manifested as total energy unboundedness from below.

Our analysis (see Supplementary Information (SI)~\cite{SI}) traces the BC instability to the ill-defined nature of $\rho_\mathrm{avg}$ as a rigid spatial average that remains fixed under density scaling $\rho_\sigma(\mathbf{r}) = \sigma^3\rho_1(\sigma\mathbf{r})$ in isolated systems.
This unphysical constant simultaneously violates the exact scaling law $T_s[\rho_\sigma] = \sigma^2T_s[\rho_1]$~\cite{85PRA-Levy-scaling} and generates negative Pauli energy $T_\theta$ through its mismatch with the physical density $\rho(\mathbf{r})$, contradicting its fundamental positivity requirement~\cite{88PRA-Levy-pauli}.
The BC instability therefore stems from this dual failure of $\rho_\mathrm{avg}$, which both breaks scaling invariance and misrepresents the true density, collectively rendering the Pauli energy and thus the total energy unbounded.
This understanding addresses a longstanding challenge in nonlocal KEDF development.

The density-dependent kernel KEDFs further enhance the accuracy by embedding spatial density dependence into $k_{\mathrm{F}}$.
The Wang-Govind-Carter (WGC)~\cite{99B-Wang-nonlocal} and Xu-Wang-Ma (XWM)~\cite{19B-Xu-nonlocal} KEDFs, for instance, replace $k_{\mathrm{F}}^0$ with a two-body Fermi wave vector $k_{\mathrm{F}}(\rho(\mathbf{r}),\rho(\mathbf{r}^{\prime}))$, enhancing bulk accuracy.
Although Taylor expansions around reference densities $\rho_{\mathrm{ref}}$ reduce computational load, sensitivity to $\rho_{\mathrm{ref}}$ choices and its ill-definition in finite systems persist.  
Alternative adaptations like the L$X$~\cite{19B-Mi-qdot} and LDAK-$X$~\cite{20B-Xu-nonlocal} series employ local density approximation (LDA), substituting $\rho_\mathrm{avg}$ with $\rho(\mathbf{r})$ in $k_{\mathrm{F}}(\mathbf{r}) = (3\pi^2\rho(\mathbf{r}))^{1/3}$. 
This {\it ad hoc} modification both breaks kernel exchange symmetry ($w(\mathbf{r},\mathbf{r}^{\prime}) \neq w(\mathbf{r}^{\prime},\mathbf{r})$) and requires cubic Hermite spline interpolations, which inflate the prefactor $m$ in their $O(mN\ln N)$ scaling.
Even the Huang-Carter (HC) KEDF~\cite{10B-Huang-nonlocal}, which circumvents reference densities and succeeds in dimers~\cite{12JCP-Xia-hc_dimer}, requires empirical tuning of parameters $\lambda$ and $\beta$ per system --- a severe limitation for general applications.  

In this work, we have established a rigorous connection between the BC instability and the ill-defined average charge density $\rho_\mathrm{avg}$ in density-independent kernels.
%
By introducing a density-functional-dependent kernel, we eliminate this instability while preserving the $O(N\ln N)$ computational scaling of the WT KEDF.
Our extended WT (ext-WT) KEDF requires no empirical parameters and achieves high accuracy for both isolated and bulk systems, representing a significant advance toward a universal KEDF.

\section{Methods}

To eliminate the BC instability, we introduce a density-functional-dependent kernel $w(k_{\mathrm{F}}[\rho],\mathbf{r}-\mathbf{r^{\prime}})$, where $k_{\mathrm{F}}[\rho] = (3\pi^2\zeta[\rho])^{1/3}$ and $\zeta[\rho]$ is a functional of charge density.
The $\zeta[\rho]$ is constructed to satisfy three fundamental requirements: recovery of the average charge density $\rho_\mathrm{avg}$ in UEG limit, proper scaling behavior $\zeta[\rho_\sigma] = \sigma^3 \zeta[\rho_1]$ under the uniform scaling $\rho_\sigma(\mathbf{r}) = \sigma^3 \rho_1(\sigma \mathbf{r})$, and magnitude comparable to the characteristic density $\rho_\mathrm{c}$ (analyzed subsequently).
While the average charge density $\rho_\mathrm{avg}$ fails for isolated systems due to its dependence on arbitrary cell volumes, our generalized $\zeta[\rho]$ formulation maintains intrinsic density dependence without external parameters.
This leads to the ext-WT KEDF that preserves both formal consistency and computational efficiency across all electronic environments.

Firstly, the ext-WT KEDF preserves the formal stability of the original WT framework near the UEG. 
As demonstrated by Blanc and Canc\`es, WT KEDF is stable near the UEG, with the external potential $V(\mathbf{r})$ slowly varied.
Although the density-functional-dependent kernel introduce additional terms to the kinetic potential and the Hessian matrix, these vanish identically in the UEG limit due to two key mechanisms: the integral identity $\int{{w(k_\mathrm{F}[\rho], \mathbf{r}-\mathbf{r^{\prime}})} {\mathrm{d}^3 \mathbf{r^\prime}}} = 0$ and the exact recovery $\zeta[\rho]\big|_{\mathrm{UEG}} = \rho_\mathrm{avg}$ (see SI~\cite{SI}).
Thus, ext-WT retains the Lindhard linear response behavior of WT for homogeneous systems.
Computationally, the  kernel’s extra terms are efficiently evaluated via Fast Fourier Transform (FFT), preserving WT’s $O(N\ln N)$ scaling (see Fig.~S1). 
This ensures minimal overhead while extending ext-WT’s applicability to non-uniform densities.

\begin{figure}[tbp]
	\centering
	\includegraphics[width=0.98\linewidth]{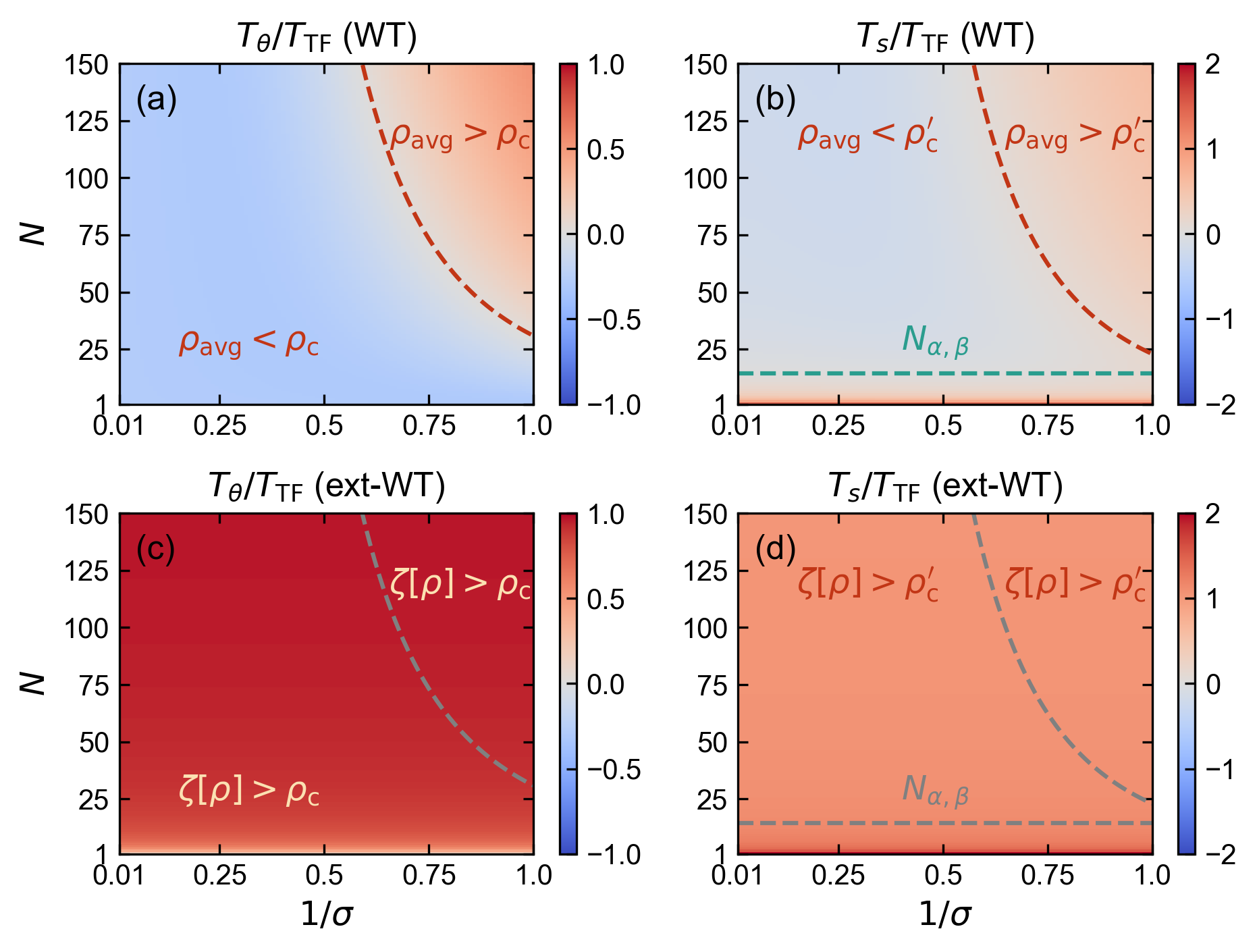}\\
	\caption{
        Pauli energy $T_{\theta}$ and non-interacting kinetic energy $T_s$ of Gaussian densities with varying electron number $N$ and scaling parameter $\sigma$.
        (a) $T_\theta / T_{\mathrm{TF}}$ and (b) $T_s / T_{\mathrm{TF}}$ computed via the WT KEDF, where $T_{\mathrm{TF}}$ is the Thomas-Fermi kinetic energy.
        The estimated boundaries $\rho_{\mathrm{avg}}=\rho_{\mathrm{c}}$ (for $T_\theta$) and $\rho_{\mathrm{avg}}=\rho_{\mathrm{c}}^{\prime}$ (for $T_s$), as determined by characteristic density $\rho_{\mathrm{c}}$ and $\rho_{\mathrm{c}}^{\prime}$, closely align with the transition between positive and negative energies. The critical particle number $N_{\alpha, \beta}$ predicted by Blanc and Canc\`es demonstrates quantitative agreement with numerical results.
        (c, d) Corresponding $T_\theta / T_{\mathrm{TF}}$ ($T_s / T_{\mathrm{TF}}$) for the ext-WT KEDF.
        The consistent satisfaction of $\zeta[\rho] > \rho_{\mathrm{c}}$ ($\zeta[\rho] > \rho_{\mathrm{c}}^{\prime}$) across all $N$ and $\sigma$ ensures non-negative energies.
        The $\sigma$-invariance of these ratios confirms strict adherence to the scaling law.
        }
        \label{fig:Gaussian}
\end{figure}

\begin{figure*}[tbp]
    \centering
    
    \begin{subfigure}{0.98\textwidth}
    \centering
    \includegraphics[width=0.99\linewidth]{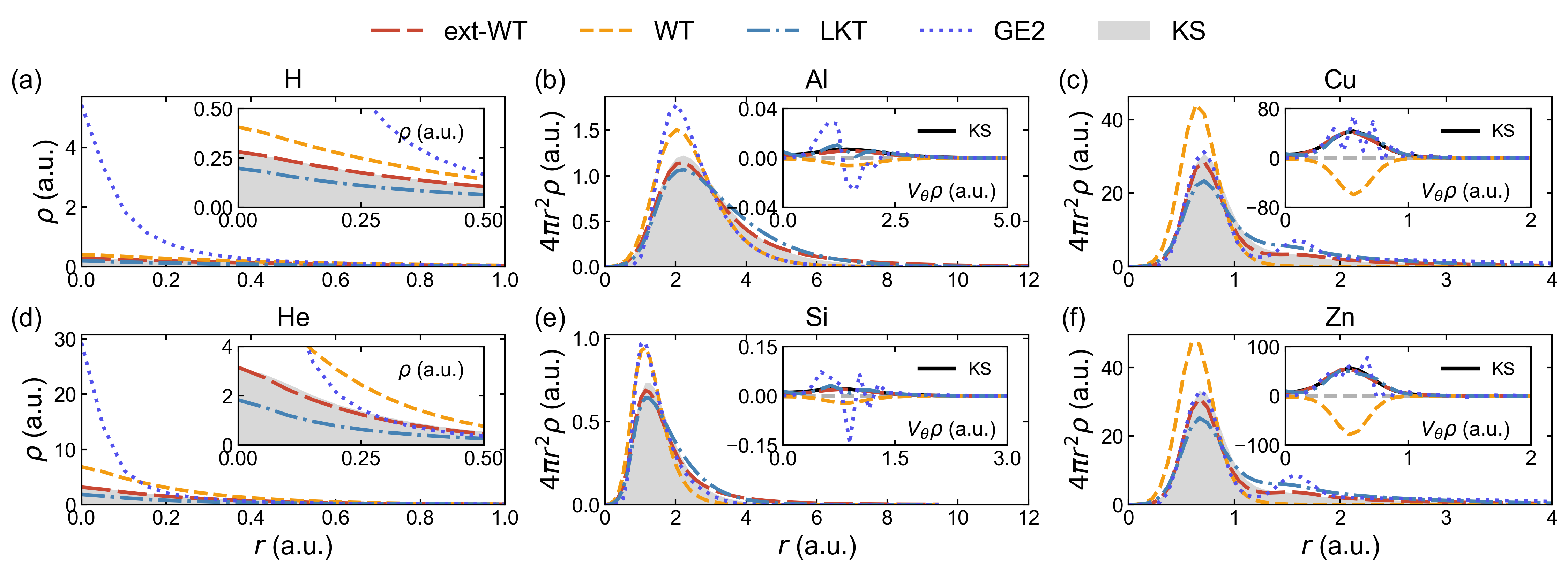}
    \label{fig:atoms}
    \end{subfigure}
    
    \caption{
    Charge density profiles for selected atoms.
    (a, d) H and He densities obtained with bare Coulomb potentials, with insets highlighting nuclear cusp behavior.
    (b, e) Al and Si using BLPS.
    (c, f) Cu and Zn employing HQLPS.
    Insets in (b, c, e, f) depict the spatial product $V_{\theta}(\mathbf{r})\rho(\mathbf{r})$.
    }
    \label{fig:Den}
\end{figure*}

\begin{figure*}[thbp]
    \centering
    
    \begin{subfigure}{0.9\textwidth}
    \centering
    \includegraphics[width=0.98\linewidth]{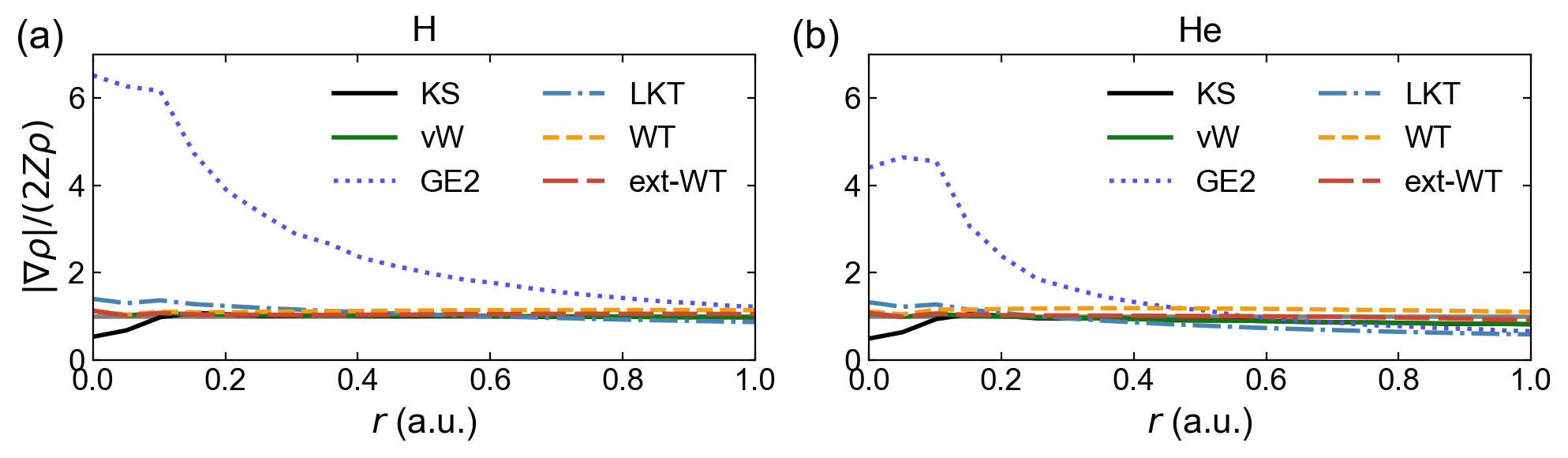}
    \end{subfigure}
    
    \caption{The Kato’s nuclear cusp condition of (a) H and (b) He, as obtained by KSDFT, and GE2, LKT, WT, and ext-WT KEDFs. The charge densities obtained by ext-WT KEDF satisfy the condition $\lim_{\mathbf{r}\to \mathbf{R}_i}{\frac{|\nabla\rho(\mathbf{r})|}{2Z_i\rho(\mathbf{r})}} \sim 1$, where $\mathbf{R}_i$ and $Z_i$ denote the nuclear coordinate and atomic numbers, respectively.}
    \label{fig:kato}
\end{figure*}

\begin{table*}[htbp]
	\centering
	\caption{
        MARE for total energies $E_{\mathrm{tot}}$ and MAE for charge densities $\rho(\mathbf{r})$, comparing results from various KEDFs to KSDFT benchmarks across 56 single-atom systems.
        Systems include 2 atoms with bare Coulomb potentials, 9 with BLPS, and 45 with HQLPS.
    }
	\begin{tabular*}{0.8\linewidth}{@{\extracolsep{\fill}} l c c c c @{}}
		\hline\hline
		MARE of $E_{\mathrm{tot}}$ (\%)       &Coulomb (2)    &BLPS (9)    &HQLPS (45)    &Total (56)\\
		\hline
            GE2     &$39.4$    &$11.0$    &$5.2$   &$7.4$\\
            LKT     &$29.0$    &$7.5$    &$6.6$   &$7.6$\\
            WT      &$26.5$    &$18.9$    &$43.2$   &$38.7$\\
            ext-WT  &\textbf{9.0}    &\textbf{2.8}    &\textbf{1.3}   &\textbf{1.8}\\
            \hline
		MAE of $\rho(\mathbf{r})$ (a.u.)       &Coulomb (2)    &BLPS (9)    &HQLPS (45)    &Total (56)\\
		\hline
            GE2     &$9.2\times10^{-5}$    &$3.8\times10^{-5}$    &$4.4\times10^{-4}$   &$3.6\times10^{-4}$\\
            LKT     &$1.3\times10^{-4}$    &$2.2\times10^{-5}$    &$4.3\times10^{-4}$   &$3.5\times10^{-4}$\\
            WT      &\boldmath{$7.5\times10^{-5}$}    &$4.0\times10^{-5}$    &$1.2\times10^{-3}$   &$1.0\times10^{-3}$\\
            ext-WT  &$8.4\times10^{-5}$    &\boldmath{$1.9\times10^{-5}$}    &\boldmath{$3.3\times10^{-4}$}   &\boldmath{$2.7\times10^{-4}$}\\
		\hline\hline
	\end{tabular*}
	\label{tab:mae}
\end{table*}

\begin{figure*}[tbp]
    \centering
    
    \begin{subfigure}{0.98\textwidth}
    \centering
    \includegraphics[width=0.99\linewidth]{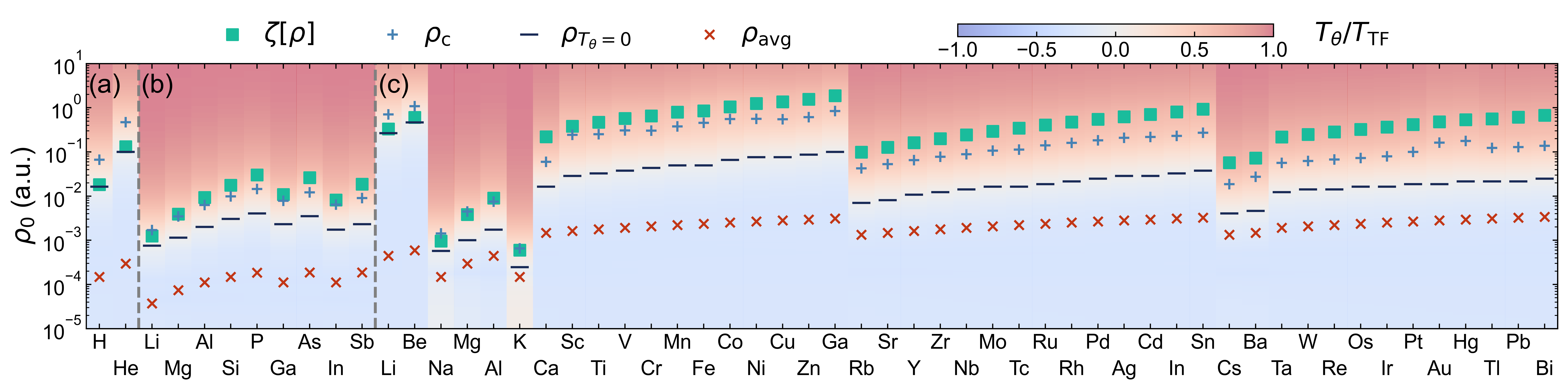}
    \label{fig:rho0}
    \end{subfigure}
    
    \caption{
    Comparison of the density functional $\zeta[\rho]$, characteristic density $\rho_{\mathrm{c}}$, and average charge density $\rho_{\mathrm{avg}}$ across single-atom systems employing (a) bare Coulomb potentials, (b) BLPS, and (c) HQLPS.
    Background color maps represent $T_\theta/T_{\mathrm{TF}}$ ratios from WT KEDF calculations across $\rho_0$ values, the dark-blue dashes specify $\rho_{T_\theta=0}$, which render $T_\theta=0$.
    All data were derive from the charge densities obtained by ext-WT KEDF.
    }
    \label{fig:Pauli_e}
\end{figure*}

Secondly, the ext-WT KEDF rigorously preserves the scaling law $T_{\mathrm{NL}}[\rho_\sigma] = \sigma^2 T_{\mathrm{NL}}[\rho_1]$ through the covariance relation $k_{\mathrm{F}}[\rho_\sigma] = \sigma k_{\mathrm{F}}[\rho_1]$.
Furthermore, it guarantees positivity of the Pauli energy through judicious construction of $\zeta[\rho]$.
For the ext-WT KEDF, we establish a characteristic density threshold which ensures $T_\theta \ge 0$:
\begin{equation}
    \begin{aligned}
        \zeta[\rho] \ge \rho_{\mathrm{c}} \equiv \left[\frac{4}{25}\frac{\int{\left|\nabla\rho^{5/6}(\mathbf{r})\right|^2 {\mathrm{d}}^3\mathbf{r}}}{T_{\mathrm{TF}}}\right]^{3/2},
    \end{aligned}
\end{equation}
which serves as a physically meaningful reference scale for $\zeta[\rho]$.
Numerical verification shows this represents a conservative bound, as $T_\theta\ge0$ persists when $\zeta[\rho]$ slightly undershoots $\rho_\mathrm{c}$.
The existence of such a scale ensures that appropriate $\zeta[\rho]$ functionals can always enforce Pauli energy positivity across arbitrary density distributions.
Additionally, a related characteristic density $\rho_\mathrm{c}^\prime$ ensuring $T_s\ge 0$ can be derived in a similar manner.
Complete derivations are provided in the SI~\cite{SI}.

The ext-WT KEDF thus adheres rigorously to the scaling law and ensures the positivity of the Pauli energy.
These advancements eliminate the BC instability while retaining the original framework’s key advantages: Lindhard response behavior near the UEG limit and $O(N\ln N)$ computational scaling.
While the preceding analysis holds for general $\zeta[\rho]$ satisfying the three design principles, practical implementation requires explicit functional forms. 
We propose
\begin{equation}
    \zeta[\rho] = \frac{\int{\rho^{\kappa + 1}(\mathbf{r}){\mathrm{d}}^3 \mathbf{r}}}{\int{\rho^{\kappa}(\mathbf{r}){\mathrm{d}}^3 \mathbf{r}}},
\end{equation}
where $\kappa=0$ recovers the conventional average density $\rho_\mathrm{avg}$ while $\kappa>0$ enforces the required constraints.
Crucially, the parameter $\kappa = \frac{1}{2(4/3)^{1/3} - 1}\approx 0.832$ is analytically determined by replacing the hydrogen (H) atom’s exact solution with a uniform electron density distribution while preserving the average nucleus-electron separation (see SI~\cite{SI}).
This ansatz provides a universally applicable KEDF requiring no empirical parameters beyond those in the original WT formulation.

\begin{figure*}[tbp]
    \centering
    
    \begin{subfigure}{0.49\textwidth}
    \centering
    \includegraphics[width=0.98\linewidth]{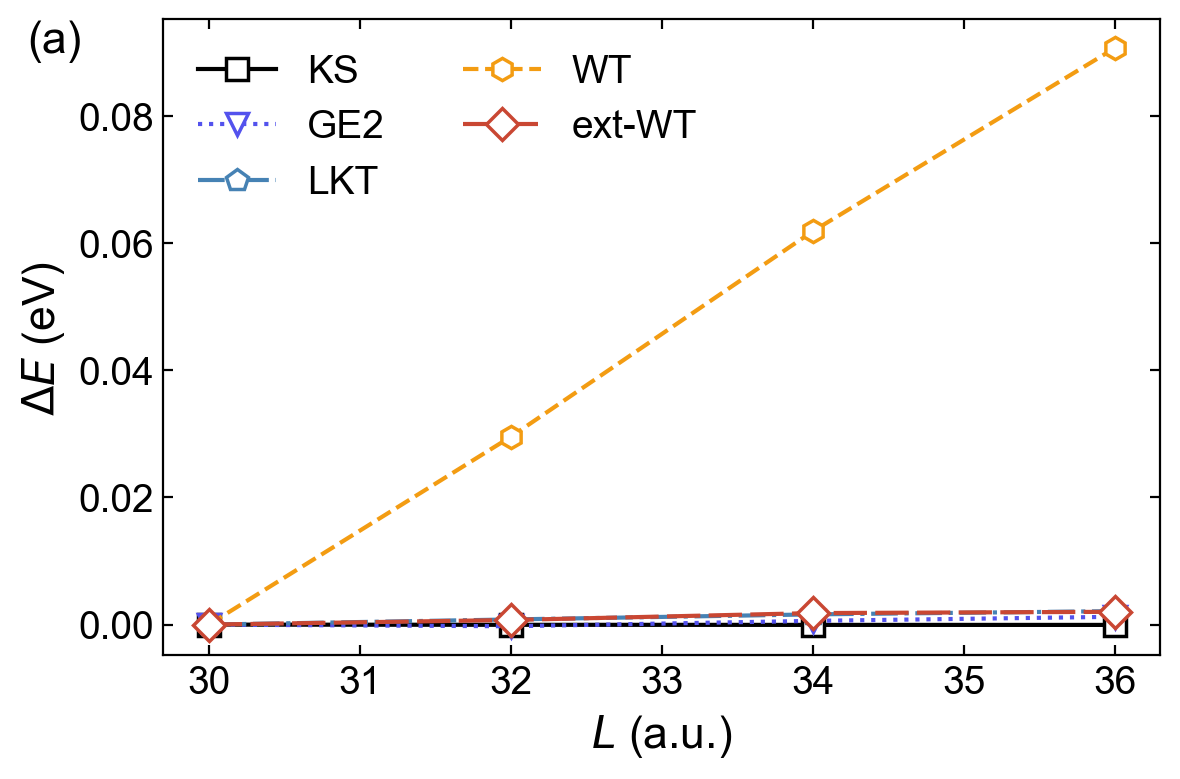}
    \end{subfigure}
    \begin{subfigure}{0.49\textwidth}
    \centering
    \includegraphics[width=0.98\linewidth]{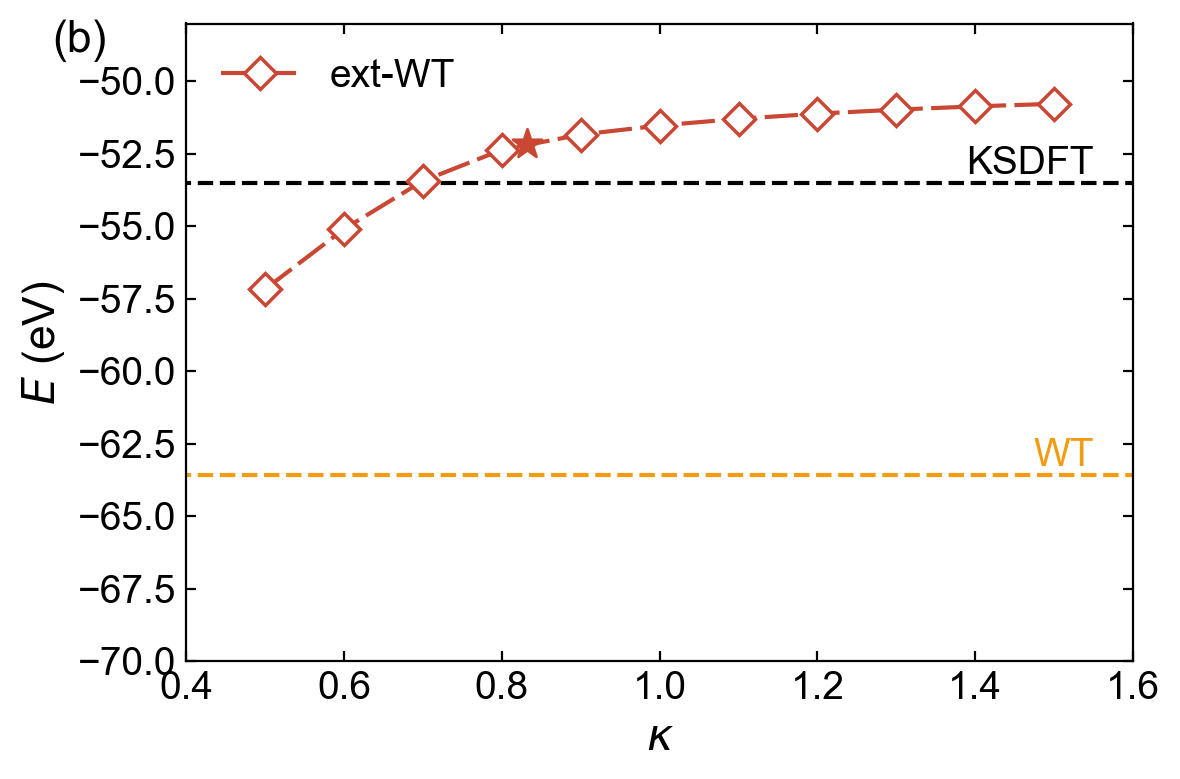}
    \end{subfigure}
    
    \caption{
    Dependence of the total energy on (a) the simulation cell size $L$ and (b) the parameter $\kappa$ for an Al atom. In (a), the relative energy change $\Delta E(L) = E(L) - E(L=30)$ is shown for KSDFT, GE2, LKT, WT, and ext-WT KEDFs at $L=$ 30, 32, 34, and 36 a.u.
    In (b), the total energy of the ext-WT KEDF is plotted across the range $\kappa=0.5$ to 1.5, with the black and yellow lines representing the corresponding results from KSDFT and the WT KEDF, respectively.
    The red star marks the adopted value $\kappa\approx0.832$.
    }
    \label{fig:L_kappa}
\end{figure*}

\begin{figure}[tbp]
	\centering
	\includegraphics[width=0.98\linewidth]{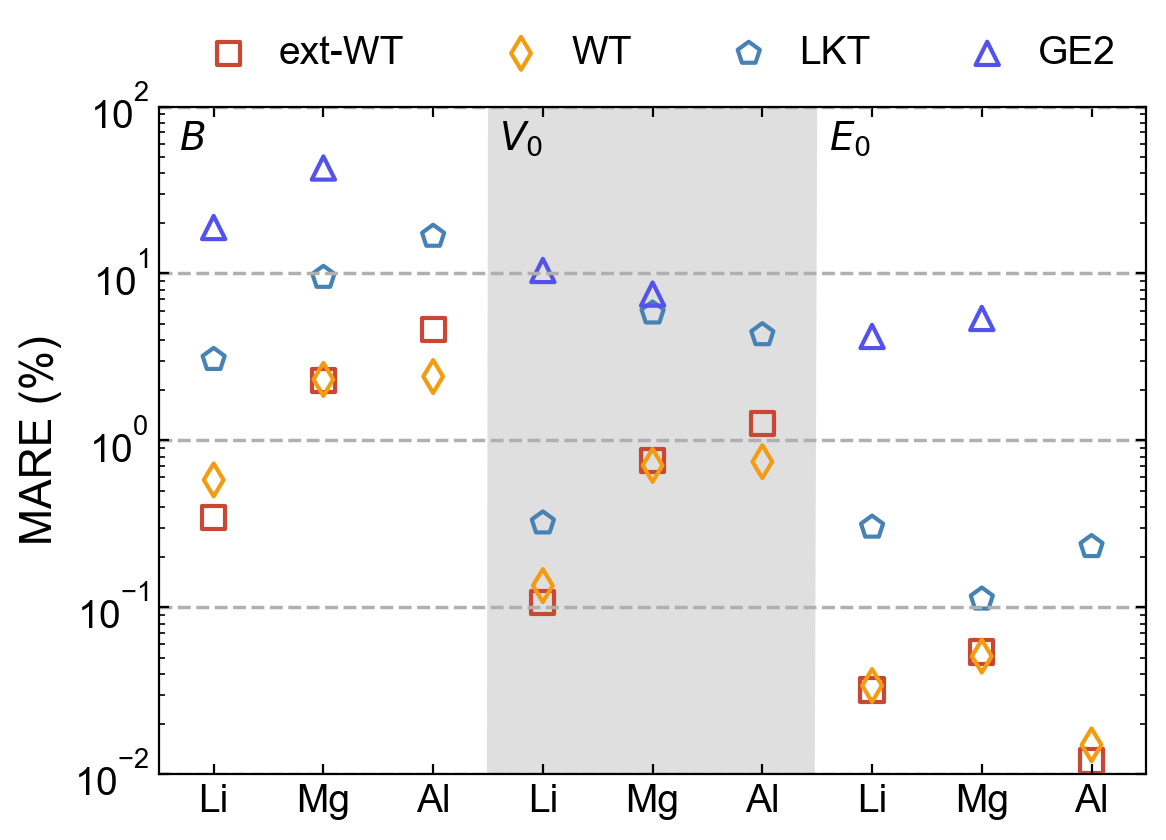}\\
	\caption{
        Mean absolute relative errors (MAREs) of bulk properties across Li, Mg, and Al systems: the bulk moduli ($B$ in $\text{GPa}$), the equilibrium volumes ($V_0$ in $\text{\AA}^3$/atom), and the equilibrium energies ($E_0$ in eV/atom), benchmarked against KSDFT with BLPS.
        Calculations span multiple crystal structures: body-centered cubic (bcc), face-centered cubic (fcc), simple cubic (sc), and cubic diamond (CD) for Li; hexagonal close-packed (hcp), fcc, bcc, and sc for Mg; fcc, hcp, bcc, and sc for Al.
        Results for Al using the GE2 KEDF are excluded due to its failure to find the equilibrium volumes.
        }
        \label{fig:Bulk}
\end{figure}

\section{Results}

To validate the elimination of the BC instability, we analyze the ext-WT and WT KEDFs using a Gaussian charge density confined within a cubic cell
\begin{equation}
    \rho_{N, r_0, L}(\mathbf{r}) = N \left(\frac{1}{\pi r_0^2}\right)^{3/2} e^{-\mathbf{r}^2/r_0^2},
\end{equation}
where $N$, $r_0$, and $L$ denote the electron number, the characteristic width ($r_0 \ll L$), and the cell length, respectively.
For $L=10$ a.u. and $r_0=1$ a.u., we scale the density as $\rho_{\sigma}(\mathbf{r}) = \sigma^3\rho_1(\sigma\mathbf{r})$ to probe energy behavior across $N$ and scaling factor $\sigma$.
Fig.~\ref{fig:Gaussian} displays the computed Pauli energy $T_\theta$ and non-interacting kinetic energy $T_s$.
As predicted by BC instability, the WT KEDF yields negative $T_\theta$ and $T_s$ at large $\sigma$, with transition boundaries aligning quantitatively with our derived characteristic densities $\rho_\mathrm{c}$ and $\rho_\mathrm{c}^\prime$.
%
%
For the ext-WT KEDF, the non-negativity of $T_\theta$ and $T_s$ is guaranteed by the satisfaction of $\zeta[\rho]\ge\rho_{\mathrm{c}}$ ($\zeta[\rho] > \rho_{\mathrm{c}}^{\prime}$) across all $N$ and $\sigma$ (see details in SI~\cite{SI}).
Crucially, the $\sigma$-invariant ratios $T_\theta / T_{\mathrm{TF}}$ and $T_s / T_{\mathrm{TF}}$ confirm strict adherence to the scaling law $T[\rho_{\sigma}]=\sigma^2T[\rho_1]$.
By simultaneously restoring Pauli energy positivity and scaling law, the ext-WT KEDF resolves the BC instability.

With the BC instability resolved, we benchmark the ext-WT KEDF against semilocal KEDFs, including the second-order gradient expansion (GE2)~\cite{57SPJETP-Kirzhnits-GE2} and Luo-Karasiev-Trickey (LKT) KEDFs~\cite{18B-Luo-semilocal}, alongside the nonlocal WT KEDF, and KSDFT for single-atom systems of 56 elements.
All calculations were performed within the ABACUS 3.8.0 packages~\cite{16CMS-Li-ABACUS, 25arXiv-Zhou-ABACUS} with Perdew-Burke-Ernzerhof (PBE)~\cite{96PRL-Perdew-PBE} exchange-correclation functional, employing three distinct potentials: bare Coulomb potential for H and helium (He), widely-used bulk-derived local pseudopotentials (BLPS)~\cite{08PCCP-Huang-BLPS} covering nine elements, and recently proposed high-quality local pseudopotentials (HQLPS)~\cite{24JCTC-Chi-HQLPS} covering all simple and transition metals.
\SL{While all-electron OFDFT calculations for heavier atoms are of theoretical interest, they are computationally prohibitive in plane-wave frameworks due to the high kinetic energy cutoffs required to describe core electrons.
Therefore, following standard practice in plane-wave OFDFT, we employ the pseudopotential approach for this study and leave broader all-electron investigations for future work.}

Figs.~\ref{fig:Den}(a) and (d) display the H and He density profiles, where the ext-WT KEDF exhibits near-quantitative agreement with KSDFT charge densities, significantly outperforming semilocal methods where GE2 shows unphysical density over-localization and LKT provides only moderate improvement.
As demonstrated in Fig.~\ref{fig:kato}, ext-WT KEDF satisfies Kato’s nuclear cusp condition $\lim_{\mathbf{r}\to \mathbf{R}_i}{\left[|\nabla\rho(\mathbf{r})| - 2Z_i\rho(\mathbf{r})\right]} = 0$, a critical feature that is notably violated by semilocal KEDFs such as GE2 and LKT.
Notably, the ext-WT KEDF yields nearly identical results to the vW KEDF, which is exact in H and He atoms, confirming its exceptional accuracy in describing core electron behavior.
%

For heavier elements such as aluminum (Al), silicon (Si), copper (Cu), and zinc (Zn), Fig.~\ref{fig:Den}(b), (e), (c), (f) present the charge densities and Pauli potential products $V_\theta(\mathbf{r})\rho(\mathbf{r})$.
While WT yields unphysical negative Pauli potentials and semilocal methods exhibit oscillatory artifacts, ext-WT accurately reproduces the positive-definite profiles of KSDFT with quantitative fidelity.
Additional results for charge densities in other tested atomic systems are provided in Figs.~S3 and S4, while the corresponding Pauli potentials are shown in Figs.~S5, S6, and S7.


Table~\ref{tab:mae} summarizes the performance of various KEDFs across 56 single-atom systems, presenting the mean absolute relative error (MARE) for total energies $E_{\mathrm{tot}}$ and the mean absolute error (MAE) for charge densities.
The ext-WT KEDF achieves the highest accuracy, yielding a total‑energy MARE of 1.8\%, which is notably lower than those of semilocal KEDFs (GE2: 7.4\%, LKT: 7.6\%) and represents a 20‑fold improvement over the original WT KEDF (38.7\%).
For charge densities, ext‑WT also exhibits the smallest MAE ($2.7\times10^{-4}$ a.u.), outperforming all other tested KEDFs.
Complete comparisons of total energies and Pauli energies are provided in SI~\cite{SI} Tables S2 and S3, respectively.
Overall, the ext‑WT KEDF substantially enhances the accuracy of the WT KEDF for single‑atom systems and surpasses the performance of conventional semilocal KEDFs.

Fig.~\ref{fig:Pauli_e} systematically compares $\zeta[\rho]$ against characteristic densities $\rho_\mathrm{c}$ and average charge densities $\rho_{\mathrm{avg}}$.
The two-order magnitude disparity between $\rho_{\mathrm{avg}}$ and $\rho_\mathrm{c}$ confirms the inadequacy of simple averaging in isolated systems.
While most $\zeta[\rho]$ values slightly exceed $\rho_\mathrm{c}$, all remain within the same order of magnitude, ensuring positive Pauli energies.
These results establish ext-WT as a reliable KEDF for these systems.

Fig.~\ref{fig:L_kappa} systematically examines the performance of the ext-WT KEDF with respect to two key aspects: the dependence of total energy on simulation cell size $L$ and the sensitivity to the parameter $\kappa$. 
As shown in Fig.~\ref{fig:L_kappa}(a), while the original WT KEDF exhibits an unphysical variation of total energy with $L$, the ext-WT KEDF remains virtually constant across different cell sizes, in close agreement with KSDFT and established semilocal KEDFs such as LKT and GE2. 
This confirms the elimination of the pathological $L$-dependence inherent in the WT KEDF.
Furthermore, Fig.~\ref{fig:L_kappa}(b) illustrates the variation of total energy with $\kappa$ for the Al atom.
We note that the adopted value $\kappa\approx 0.832$ is not optimized to minimize the energy error for Al.
While system-specific $\kappa$ optimization could yield higher numerical accuracy, we retain a physically derived $\kappa$ to ensure consistency and universality across systems.

Fig.~\ref{fig:Bulk} illustrates the performance of the ext-WT KEDF in predicting bulk properties (bulk moduli, equilibrium volumes, and energies) for simple metals (Li, Mg, Al), validating the UEG-proximity stability inherited from the WT KEDF.
Benchmarking against KSDFT reveals that the ext-WT KEDF preserves the exceptional accuracy of the WT method while surpassing semilocal KEDFs by over an order of magnitude in MARE.
This dual capability enables the ext-WT KEDF to achieve high accuracy for both isolated and bulk systems, establishing a unified framework for heterogeneous material simulations.

\section{Conclusions}

In summary, we systematically demonstrate that the ill-defined average charge density $\rho_{\mathrm{avg}}$ in isolated systems induces the BC instability by violating scaling laws and disrupting Pauli energy positivity.
By introducing a density-functional-dependent kernel, we eliminate this instability while preserving the computational efficiency and UEG accuracy of the original WT framework.
In 56 single-atom calculations, ext-WT reduces total energy errors by an order of magnitude compared to WT KEDF, while maintaining the lowest charge density errors among all tested functionals. 
This work establishes a generalizable strategy for designing nonlocal KEDFs that bridge bulk and isolated systems, addressing a longstanding limitation in OFDFT.

\section{Acknowledgements}

The work of L.S. and M.C. was supported by
the National Natural Science Foundation of China Excellence Research Group Program No.~12588301 and the National Key R$\&$D Program of China under Grant No. 2025YFB3003603.
The numerical simulations were performed on the High-Performance Computing Platform of CAPT.
\SL{This work was partially supported by PKU Kunpeng\&Ascend Center of Excellence.}

\bibliography{ext-WT}


\end{document}